\documentclass[10pt,letterpaper]{article}
\usepackage{opex3}
\begin{document}
\title{Coherent Population Trapping in Diamond N-V Centers at Zero Magnetic Field}
\author{Charles Santori$^1$, David Fattal$^1$, Sean M. Spillane$^1$, Marco Fiorentino$^1$, Raymond G. Beausoleil$^{1}$, Andrew D. Greentree$^{2,4}$, Paolo Olivero$^{4}$, Martin Draganski$^{5}$, James R. Rabeau$^{4}$, Patrick Reichart$^{4}$, Brant C. Gibson$^{3,4}$, Sergey Rubanov$^{2,4}$, David N. Jamieson$^{2,4}$, Steven Prawer$^{2,4}$}

\address{$^1$Hewlett-Packard Laboratories,
             1501 Page Mill Rd., Palo Alto, CA 94304}
\address{$^2$Centre for Quantum Computer Technology}
\address{$^3$Quantum Communications Victoria}
\address{$^4$School of Physics, The University of Melbourne,
              Melbourne, Victoria 3010, Australia}
\address{$^5$Applied Physics, RMIT University,
             GPO Box 2476V, Melbourne, Victoria 3001, Australia}

\email{charles.santori@hp.com}

\homepage{http://www.hpl.hp.com/research/qsr/}


\begin{abstract}
Coherent population trapping at zero magnetic field was observed for
nitrogen-vacancy centers in diamond under optical excitation. This
was measured as a reduction in photoluminescence when the detuning
between two excitation lasers matched the 2.88~GHz crystal-field
splitting of the color center ground states.  This behavior is
highly sensitive to strain, which modifies the excited states, and
was unexpected following recent experiments
demonstrating optical readout of single nitrogen-vacancy electron
spins based on cycling transitions. These results demonstrate for
the first time that three-level Lambda configurations suitable for
proposed quantum information applications can be realized
simultaneously for all four orientations of nitrogen-vacancy centers
at zero magnetic field.
\end{abstract}




Impurity spins in solids are appealing as physical systems for
quantum information processing, combining long decoherence times
with the possibility of large-scale integration based on
semiconductor processing technology~\cite{kane98}. Of the various
impurities offering access to individual electron spins, one of the
most promising is the nitrogen-vacancy (N-V) center in diamond,
which consists of a substitutional nitrogen atom next to a carbon
vacancy. Electron spin coherence times up to $58 \, \mu {\rm s}$
have been measured at room temperature~\cite{kennedy03}, and optical
readout of the electronic spin state of a single N-V center has been
demonstrated~\cite{jelezko02}. Controlled coupling between
electronic and nuclear spins has been
reported~\cite{wilson03,jelezko04}, offering potentially much longer
storage times, and controlled coupling with nearby nitrogen
impurities is also possible~\cite{epstein05}. Single-photon
generation at room temperature~\cite{kurtsiefer00,beveratos02} has
established the potential of N-V centers for quantum communication.
Based on such experimental results, schemes for quantum memories and
repeaters~\cite{childress05} and quantum computation using electron
spins~\cite{shahriar02,nizovtsev05,greentree06} have been proposed.

Most experiments conducted on N-V centers to date have relied on
microwave fields to manipulate the electron spins and have used
optical fields mainly for readout.  All-optical control would allow
for spatially selective addressing of single or a few N-V
centers~\cite{shahriar02} and could be used in nonlinear-optical
devices based on electromagnetically-induced transparency
(EIT)~\cite{schmidt96,munro03}. It is generally believed that the
negatively charged N-V center has optical transitions between the
$^3A$ ground states and $^3E$ excited states at zero magnetic field
that are almost perfectly spin-preserving.  This situation would be
advantageous for non-destructive readout of electron spins through
photoluminescence detection, but---if this is the general case---poses a
problem for all-optical control of the electron spins.  All-optical
control requires a $\Lambda$ system, which has two ground-state
levels connected to a common excited level by optical transitions.
One solution is to work close to an avoided crossing of the $m_s =
0$ and $m_s = -1$ ground states that occurs at a particular magnetic
field, as described in Ref.~\cite{hemmer01}, where EIT was first
reported in this system. However, previous spectral hole-burning
studies~\cite{reddy87,manson94} suggest that the optical transitions
are not always spin-preserving, even at zero magnetic field. In
these experiments, anti-hole features (increases in
photoluminescence) were observed when two excitation lasers were
separated by $\pm 2.88 \, {\rm GHz}$, the crystal-field splitting
energy between the $m_s = 0$ and the degenerate $m_s = \pm 1$ ground
states.  These features can only appear if one or more excited
states couples to both the $m_s = 0$ and $m_s = \pm 1$ levels. Thus
there is an apparent contradiction between the appearance of these
anti-hole features and the more recent experiments demonstrating
optical readout based on cycling transitions. A possible clue in
Ref.~\cite{manson94} is that the anti-hole structure showed a
dependence on the center frequency, which could be interpreted as a
strain effect. Nevertheless, the excited-state structure remains
poorly understood, and it is important therefore to clarify the
conditions necessary to obtain spin-nonpreserving transitions as
needed for devices based on EIT and all-optical control of electron
spins.

Here, we report spectral hole-burning experiments on a sample with
little inhomogeneous broadening and a range of strain conditions
obtained by ion implantation.  We find that it is possible to
realize a $\Lambda$ system even at zero magnetic field as confirmed
through the observation of coherent population
trapping~\cite{arimondo76,whitley76}, the same mechanism responsible
for EIT. The results confirm that strain has a strong effect on the
excited-state fine structure~\cite{manson94} and suggest that N-V
centers can be engineered to have either spin-preserving or
spin-nonpreserving transitions as needed for a specific
application.

\begin{figure}[htp]
\centering\includegraphics[]{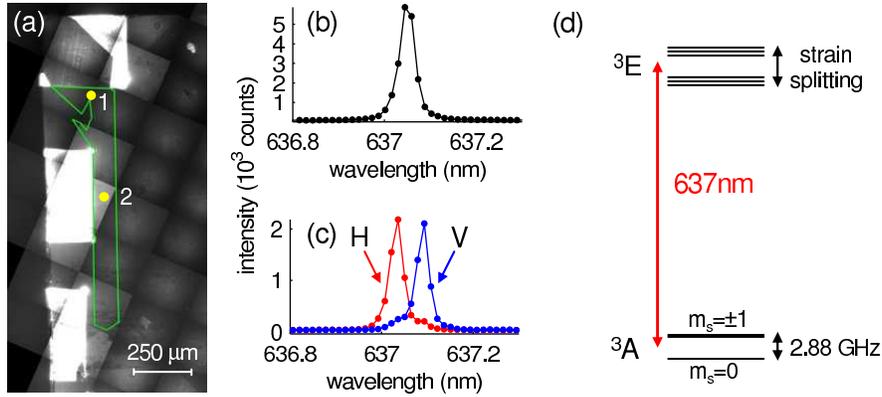}
\caption{\label{fig1} (a) Composite photoluminescence (PL) image
of a portion of the sample.
The narrow-linewidth sector described in the text is outlined in green.
(b) PL spectrum showing the zero-phonon line for negatively charged
N-V centers, measured at
location 1 in the image.  (c) Polarized PL spectra from location 2.
``H'' and ``V'' are for excitation and collection polarizations oriented
horizontally and vertically, respectively, compared with the image.
(d) Schematic energy level diagram.}
\end{figure}
The sample used for this study was a type Ib high-pressure,
high-temperature (HPHT)-grown diamond crystal obtained originally
from Sumitomo.  Figure~1(a) shows a composite photoluminescence
image obtained under $532\,{\rm nm}$ illumination. The bright squares
were implanted with $2\,{\rm MeV}$ He$^+$ ions at a concentration
of $\sim 10^{16}~\mathrm{cm}^{-2}$.
The ion implantation breaks bonds~\cite{hunn95}.
Above a certain critical damage level,
approximately $10^{22} \, \mathrm{cm}^{-3}$ vacancy
concentration~\cite{uzan_saguy95,orwa99},
upon annealing the damaged material will relax to $sp^2$ bonded
carbon.  This graphitized material has a lower density than
diamond which causes volume expansion and in turn produces strain
in the surrounding regions.
Annealing was performed at a temperature of
$1400~^{\circ}{\rm C}$ for 15 minutes in vacuum.
Although nitrogen occurs throughout
the crystal, most of the luminescence was observed to originate from
close to the surface.  Another important feature of this sample
is the presence of distinct sectors.  Growth sectors are regions that
have grown from particular faces of the original seed crystal.
Because different faces grow at different rates and incorporate
impurities at different levels, the sample shows marked
discontinuities at sector boundaries.  In
two particular growth sectors, one of which is outlined in
Fig.~1(a), the inhomogeneous linewidth of the zero-phonon
line (ZPL) for negatively charged N-V centers
at $637~{\rm nm}$ was approximately
$10-20 \, {\rm GHz}$, exceptionally narrow
compared with the more typically observed value of
$\sim 750~\mathrm{GHz}$ \cite{Zaitsev}.
Figs.~1(b,c) show low-temperature photoluminescence spectra
from two locations, measured at $10\,{\rm K}$
using $532\,{\rm nm}$ excitation.
At location 1 a single unpolarized peak was observed,
while at location 2 the ZPL was split into two
orthogonally polarized components.  We attribute the splitting at location 2
to strain induced by the nearby implanted region, the edge of
which is approximately parallel to one of
the cubic crystal axes.  By comparison with the stress
measurements in Ref.~\cite{davies76}, we estimate that
the $44.5 \, {\rm GHz}$ splitting corresponds to
an equivalent uniaxial stress of $43 \, {\rm MPa}$.
%
%
N-V centers can be oriented along four possible directions
($[111]$, {\it etc.}), but the splitting should be
the same in each case for strain along $[100]$.
A schematic energy-level diagram,
based on Refs.~\cite{martin99,manson06},
is shown in Fig.~1(d).

\begin{figure}[htb]
\centering\includegraphics[]{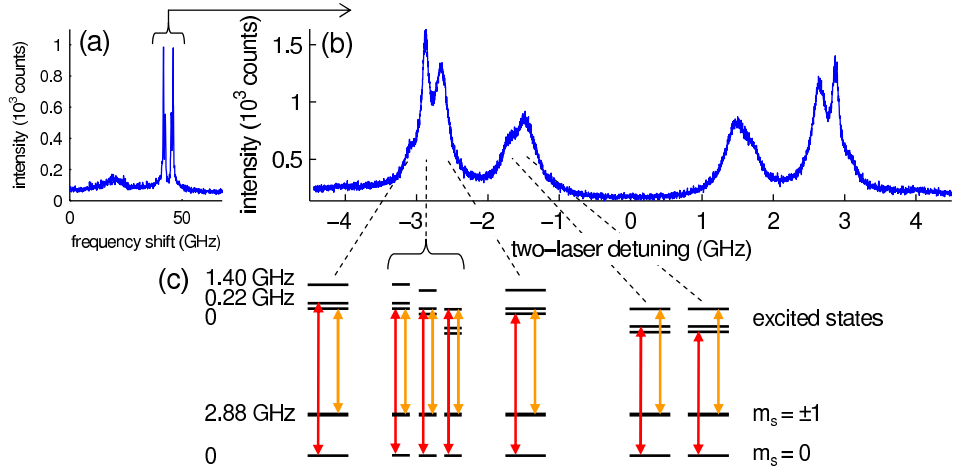}
\caption{\label{fig2}
(a) Two-laser scan at location 2, with the fixed laser tuned
to the center of the long-wavelength peak in Fig.~1c, producing
the prominent anti-hole features.  Both lasers were polarized
along ``V.'' (b) Higher-resolution scan
of the anti-hole features.  (c) Energy level diagrams explaining the
individual peaks.  The orange (red) arrows represent transitions
driven by the fixed (scanning) laser.}
\end{figure}
In the two-laser experiments, the zero-phonon line
was excited on resonance using two continuous-wave,
tunable external-cavity diode lasers.
One of the lasers (laser 2) was held fixed in frequency while
the other (laser 1) was scanned over a range of up
to $80 \, {\rm GHz}$.  The scanning rate was calibrated
by means of a Fabry-Perot cavity.
The laser stability was approximately
$1 \, {\rm MHz}$ over $1 \, {\rm s}$ and $10 \, {\rm MHz}$
over a period of 10 minutes.
A weak repump laser operating at $532 \, {\rm nm}$ was also
applied continuously to re-activate N-V centers that
bleach after many excitation cycles, an effect that is
due perhaps to photo-ionization~\cite{manson05}.
The lasers were focused by a 0.6 numerical-aperture
microscope objective through the cryostat window onto
the sample surface with a spot size of approximately
$3-4\, \mu{\rm m}$.
The resulting photoluminescence (PL) was collected by the
same objective, filtered to remove laser light, and
sent to an avalanche-photodiode photon counter
(Perkin-Elmer SPCM).  A bandpass filter
selected a $40\,{\rm nm}$ bandwidth centered at
$700\,{\rm nm}$ for detection of emission into
the broad phonon sidebands.  This provided a
measure of the excited-state population.
For most of the data shown below, laser 1
was scanned at a repetition rate
of $1\,{\rm Hz}$, and the results from many scans
were summed.

Results from typical two-laser measurements performed at
low excitation power ($0.5 \, \mu{\rm W}$) at location~2
are shown in Fig.~2.  With one laser fixed on resonance
with the long-wavelength component of the ZPL, the frequency
of the second laser was scanned across the entire ZPL structure.
In Fig.~2a, a pattern of peaks appears prominently around
the frequency of the fixed laser, and these peaks are shown
with higher resolution in Fig.~2b.
The pattern of peaks is associated with the energy level
structure and transition strengths at this particular location.
As explained in Refs.~\cite{reddy87,manson94,martin99},
photoluminescence peaks occur at two-laser detunings for which
each laser is resonant with a transition involving a different
ground state.  In these situations,
the N-V center is excited from either ground state,
producing continual photoluminescence.  If only one ground state is
excited, the population is driven to the other ground state
(optical pumping), and photoluminescence is suppressed.

The observed two-laser spectrum can be explained by an energy-level diagram
with three excited states, shown in Fig.~2c, where the
energy spacing between the ground and excited states is
allowed to vary due to inhomogeneous broadening, but the
level spacings within the ground and excited-state manifolds
are constant.  It is known that the ground states, formed
from an orbital singlet, are relatively insensitive to
stain~\cite{vanoort91}, while the
excited states, formed from an orbital doublet, are highly
sensitive to strain~\cite{manson94}.
In the excited states, strain lowers the symmetry, causing
first a splitting of the orbital doublet~\cite{davies76}, and
then a modification of the spin sublevels through
spin-orbit coupling~\cite{martin99,manson06}.
The peak positions in Fig.~2b thus indicate
the excited-state fine structure.
We observed that the peak positions and intensities
in the two-laser spectra depend strongly on position
relative to the edge of the heavily implanted region,
from which we can infer a strain dependence.
However, the peaks at detunings of
$\pm 2.88 \pm 0.01 \, {\rm GHz}$,
matching the crystal-field splitting of the ground
states, do not shift measurably with strain, and only their
height changes.  At these detunings the two lasers
couple two ground states to the same excited state, and the
presence of these peaks suggests that a $\Lambda$ system
with allowed spin-nonpreserving transitions is present.

\begin{figure}[htb]
\centering\includegraphics[]{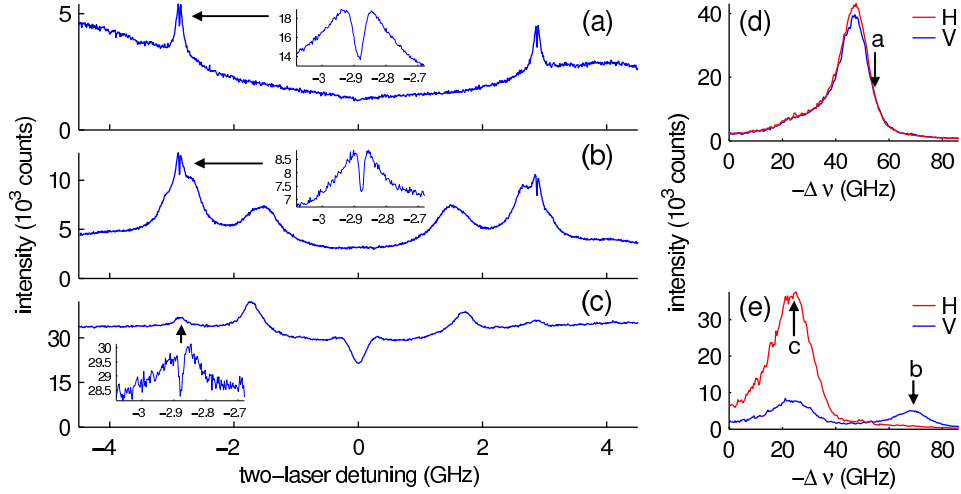}
\caption{\label{fig3}
(a,b,c) Two-laser scans showing anti-hole features and
coherent population trapping effect at sample
location 1 (a) and at location 2, measuring the long-wavelength (b)
and short-wavelength (c) components of the zero-phonon line.
Insets: higher-resolution scans of the
$-2.88 \, {\rm GHz}$ features.  Excitation powers:
$5 \, \mu{\rm W}$ for the $637 \, {\rm nm}$ lasers and
$2 \, \mu{\rm W}$ for the repump.
(d,e) Single-laser scans at locations 1 and 2,
respectively.  Arrows indicate spectral
positions studied in (a,b,c).}
\end{figure}
Observation of coherent population trapping requires only
increasing the excitation power.  Figure~3 shows
two-laser spectra measured with $5 \, \mu{\rm W}$ excitation
power at location~1 (Fig.~3a) and location~2 (Fig.~3b,c).
The single-laser spectra in Fig.~3(d,e) indicate the corresponding
spectral positions within the inhomogeneously broadened ZPLs.
The narrow dips appearing within the $2.88 \, {\rm GHz}$
anti-holes are the signature of coherent population trapping.
Close to two-photon resonance, a ``dark state''
can form, which is a coherent superposition of
ground states that has no net transition matrix
element into an excited state due to destructive
quantum interference~\cite{arimondo76,whitley76}.
This feature only appears when the transition
Rabi frequencies are large compared with the
relevant decoherence rates.  Observation of this feature
demonstrates unambiguously that we can find $\Lambda$ systems
with allowed spin-nonpreserving transitions.
At location 1, where no strain splitting is resolved,
the effect was only observed on the long-wavelength side of
the ZPL.  On the short wavelength side, the anti-hole
pattern was different, and the $2.88 \, {\rm GHz}$ anti-holes
were almost absent.
The random fluctuations in strain or
electric field responsible for the inhomogeneous broadening
apparently has a strong effect on the excited-state fine structure and
optical transitions~\cite{manson94}.
The only distinct anti-hole at this location is the
$2.88 \, {\rm GHz}$ feature associated with the
ground-state splitting, suggesting a random variation in
the excited-state structure.
At location 2, where the anti-hole structure is clearly
resolved, the $2.88 \, {\rm GHz}$ anti-holes are much more
prominent in the long-wavelength ZPL component, and the coherent
population trapping feature was easily observed there.
For the short-wavelength ZPL component, these anti-hole
features are much weaker, suggesting that the optical
transitions are nearly spin-preserving, but the coherent population
trapping feature could still be observed.

\begin{figure}[htb]
\centering\includegraphics[]{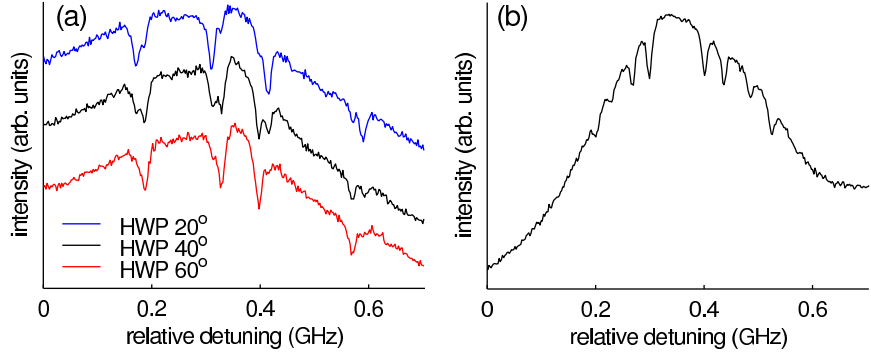}
\caption{\label{fig4} Two-laser measurements with
a weak magnetic field applied to lift the $m=\pm 1$
degeneracy.  (a) Measured at location 1.  Half-wave-plate
(HWP) angle $40^\circ$ corresponds to
polarization approximately along $[010]$.
(b) Measured at location 2 for polarization along $[010] (V)$.
All plots are scaled and shifted for clarity.  The measured
intensity decreases by approximately 8\% for the dips in (a) and
4\% for the dips in (b).
}
\end{figure}
At zero magnetic field it is possible for all
four orientations of N-V centers to show
coherent population trapping behavior.
To determine the orientation dependence, we performed
two-laser measurements with a weak magnetic
field ($\sim 100 \, {\rm Gauss}$) applied
to lift the degeneracy of the $m_s=\pm 1$ ground states.
The magnetic field direction was adjusted so that each
of the four orientations had a different
energy splitting.  As shown in Fig.~4, the photoluminescence
dip on two-photon resonance splits into a maximum of
eight components corresponding to the four orientations.
For the low-strain region (location~1), the
dips are divided into two groups responding to
polarizations along $[110]$ and $[1\bar{1}0]$.
We expect that one polarization primarily excites N-V centers
oriented along $[111]$ and $[\bar{1}\bar{1}1]$, while
the other polarization primarily excites centers
oriented along $[1\bar{1}\bar{1}]$ and $[\bar{1}1\bar{1}]$.
For polarization along $[010]$, all four
orientations produce photoluminescence dips.  For the long-wavelength
peak at location~2, eight dips appear
for $[010]$ polarization, showing that all four orientations
can participate even in the strained case.  The possibility of obtaining
equivalent $\Lambda$ systems for all four
orientations is an advantage
over the anti-crossing scheme in Ref.~\cite{hemmer01},
where only one orientation participated.

Finally, we show that the shape and excitation power
dependence of the $2.88\,{\rm GHz}$ peaks can be explained by
a theoretical model that takes into account the
inhomogeneous broadening of the ground-to-excited-state
transitions.
The power dependence measured at location 1 is shown in
Fig.~5. At low power, a shallow dip approximately $10\, {\rm MHz}$
wide appears which then deepens and broadens with increasing power.
To fit these data, we use a simplified model with three
levels in a $\Lambda$ configuration. The two ground states represent
the $m=0$ ($|1\rangle$) and a superposition of the
$m=\pm 1$ ($|2\rangle$) levels of the ground state manifold,
with a fixed frequency
difference of $2.88\, {\rm GHz}$. The excited state
($|3\rangle$) is given a variable frequency
following a normal (gaussian) distribution ($\sigma = 10 \, {\rm GHz}$).
The model includes all possible population relaxation and dephasing
terms under the simplifying assumption
that the relaxation rates and transition strengths
are the same for the two lower levels.
We first find the steady-state density matrix for a
single center and then average over
the inhomogeneous distribution. The fitted curves, shown in Fig.~5,
are proportional to the steady-state population of the excited state,
plus a linearly sloped background added as an additional fitting
parameter representing contributions from other subsets of levels.
The theoretical curves match the data quite well, confirming that the observed
behavior can be explained in terms of an inhomogeneous
distribution of three-level systems.
From the fits we estimate that the pure dephasing rate between the
excited and ground states varies in the range
$\gamma_{13} / 2\pi = 2-11 \, \mathrm{MHz}$ over the range
of excitation powers used, and this
should be compared with the radiative linewidth of
$13.4 \, {\rm MHz}$ used in the fits.
Most importantly, the fits can determine the value of the
ground-state decoherence parameter $\gamma_{12}$,
since this parameter is closely related to the shape of the dip
within the anti-hole feature.  We find
$\gamma_{12} / 2\pi = 3.5-7 \, \mathrm{MHz}$.
Laser instability can contribute to the
decoherence, but initial measurements using an electro-optic
modulator in place of a
second laser suggest that the true $\gamma_{12}$ is not much smaller
than these values for this sample.  Possible sources of
ground-state broadening include
inhomogeneous variation of the ground-state crystal-field
splitting~\cite{vanoort91},
stray magnetic fields which can affect the four orientations differently,
interactions with $N$ impurities~\cite{epstein05}
and hyperfine coupling with $^{14}N$ and $^{13}C$ nuclei.
The much longer coherence lifetime
reported in Ref.~\cite{kennedy03} was obtained using
spin-echo techniques to remove the effect of inhomogeneous
broadening.
\begin{figure}[htb]
\centering\includegraphics[]{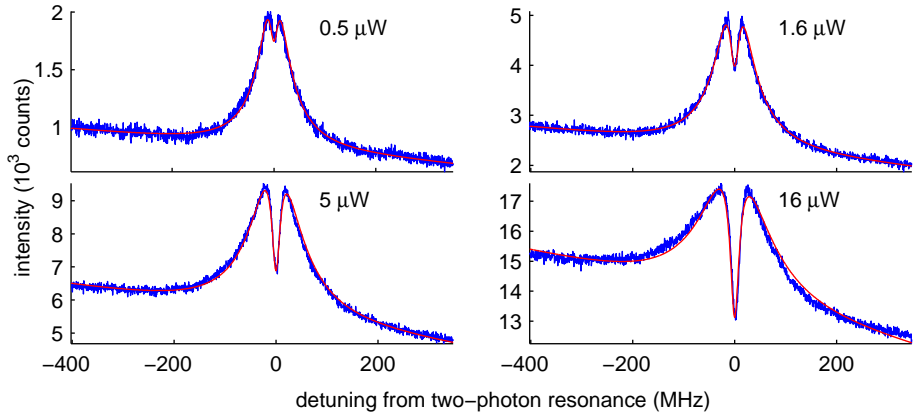}
\caption{\label{fig5} Coherent population trapping at
various excitation powers at location 1
(same as in Fig. 3a).  blue:data, red:fit.
The indicated powers correspond to both $637\,{\rm nm}$ lasers.
For excitation powers
0.5, 1.6, 5, and $16 \, \mu{\rm W}$ the fits used
Rabi frequencies (same for both transitions):
10.1, 14.2, 17.9, and $22.0 \, {\rm MHz}$;
effective population decay rates between ground states:
0.11, 0.20, 0.35, and $0.83 \, {\rm MHz}$;
excited-state decoherence:
2.4, 3.0, 6.3, and $11.4 \, {\rm MHz}$;
ground-state decoherence:
7.0, 5.8, 3.6, and $3.8 \, {\rm MHz}$.
}
\end{figure}

The coherent population trapping effects reported here
suggest that EIT and all-optical spin manipulation
should be possible with N-V centers even at zero magnetic field.
A small amount of strain,
which can be introduced through ion implantation
or other fabrication methods~\cite{olivero2005},
can determine whether spin-nonpreserving optical
transitions are allowed.
For strain along $[100]$,
all four orientations can produce a
similar $\Lambda$ system.
Therefore N-V centers have a flexibility
in their excited-state level structure that makes
them suitable either for single-spin readout through
photoluminescence detection or for optical devices based on
EIT and Raman transitions.

\subsection*{Acknowledgments}
We thank S. Shahriar and P. Hemmer for helpful
discussions.
The work performed at HP Laboratories in Palo Alto has been
supported in part by DARPA contract no.\ FA9550-05-C-0017. Quantum
Communications Victoria is supported by the State Government of
Victoria's Science, Technology and Innovation Initiative -
Infrastructure Grants Program. BCG is proudly supported by the {\it
International Science Linkages} program established under the
Australian Government's innovation statement {\it Backing
Australia's Ability}. This work was supported by the DEST,
Australian Research Council, the Australian government and by the US
National Security Agency (NSA), Advanced Research and Development
Activity (ARDA) and the Army Research Office (ARO) under contracts
W911NF-04-1-0290 and W911NF-05-1-0284.


\begin{thebibliography}{99}

\bibitem{kane98}
B. E. Kane,
``A silicon-based nuclear spin quantum computer,''
Nature (London) {\bf 393,} 133--137 (1998).


\bibitem{kennedy03}
T. A. Kennedy, J. S. Colton, J. E. Butler, R. C. Linares and P. J. Doering,
``Long coherence times at 300 K for nitrogen-vacancy center spins in diamond grown by chemical vapor deposition,''
Appl. Phys. Lett. {\bf 83,} 4190--4192 (2003).


\bibitem{jelezko02}
F. Jelezko, I. Popa, A. Gruber, C. Tietz, J. Wrachtrup, A. Nizovtsev, and S. Kilin,
``Single spin states in a defect center resolved by optical spectroscopy,''
Appl. Phys. Lett. {\bf 81,} 2160--2162 (2002).


\bibitem{wilson03}
E. A. Wilson, N. B. Manson, and C. Wei,
``Perturbing an electromagnetic induced transparency within an inhomogeneously
broadened transition,''
Phys. Rev. A {\bf 67,} 023812 (2003).


\bibitem{jelezko04}
F. Jelezko, T. Gaebel, I. Popa, M. Domhan, A. Gruber, and J. Wrachtrup,
``Observation of Coherent Oscillation of a Single Nuclear Spin and Realization of a Two-Qubit Conditional Quantum Gate,''
Phys. Rev. Lett. {\bf 93,} 130501 (2004).


\bibitem{epstein05}
R. J. Epstein, F. M. Mendoza, Y. K. Kato, and D. D. Awschalom,
``Anisotropic interactions of a single spin and dark-spin spectroscopy in diamond,''
Nature Physics {\bf 1,} 94--98 (2005).


\bibitem{kurtsiefer00}
C.~Kurtsiefer, S.~Mayer, P.~Zarda, and H.~Weinfurter,
``Stable solid-state source of single photons,''
Phys. Rev. Lett. {\bf 85,} 290--293 (2000).


\bibitem{beveratos02}
A.~Beveratos, S.~K\"{u}hn, R.~Brouri, T.~Gacoin, J.-P.~Poizat, and P.~Grangier,
``Room temperature stable single-photon source,''
Eur. Phys. J. D {\bf 18,} 191--196 (2002).


\bibitem{childress05}
L. Childress, J. M. Taylor, A. S. Sorensen, and M. D. Lukin,
``Fault-tolerant quantum repeaters with minimal physical resources and implementations based on single-photon emitters,''
Phys. Rev. A {\bf 72,} 052330 (2005).


\bibitem{shahriar02}
M. S. Shahriar, P. R. Hemmer, S. Lloyd, P. S. Bhatia, and A. E. Craig,
``Solid-state quantum computing using spectral holes,''
Phys. Rev. A {\bf 66,} 032301 (2002).


\bibitem{nizovtsev05}
A. P. Nizovtsev, S. Ya. Kilin, F. Jelezko, T. Gaebal, I. Popa, A. Gruber, and J. Wrachtrup,
``A quantum computer based on NV centers in diamond: optically detected nutations of single electron and nuclear spins,''
Opt. Spectrosc. {\bf 99,} 233--244 (2005).


\bibitem{greentree06}
A. D. Greentree,
P. Olivero, M. Draganski,
E. Trajkov, J. R. Rabeau, P. Reichart, B. C. Gibson, S. Rubanov,
S. T. Huntington, D. N. Jamieson, and S. Prawer,
``Critical components
for diamond-based quantum coherent devices,''
J. Phys.: Condens. Matter {\bf 18} S825--S842 (2006).


\bibitem{schmidt96}
H. Schmidt and A. Imamoglu,
``Giant Kerr nonlinearities obtained by electromagnetically induced transparency,''
Opt. Lett. {\bf 21,} 1936--1938 (1996).


\bibitem{munro03}
W. J. Munro, K. Nemoto, R. G. Beausoleil, and T. P. Spiller,
``High-efficiency quantum nondemolition single-photon-number-resolving detector,''
Phys. Rev. A {\bf 71,} 033819 (2005).


\bibitem{hemmer01}
P. R. Hemmer, A. V. Turukhin, M. S. Shahriar, and J. A. Musser,
``Raman-excited spin coherences in nitrogen-vacancy color centers in diamond,''
Opt. Lett. {\bf 26,} 361--363 (2001).


\bibitem{reddy87}
N. R. S. Reddy, N. B. Manson, and E. R. Krausz,
``Two-laser spectral hole burning in a colour centre in diamond,''
J. Lumin. {\bf 38,} 46--47 (1987).


\bibitem{manson94}
N. B. Manson and C. Wei,
``Transient hole burning in N-V center in diamond,''
J. Lumin. {\bf 58,} 158--160 (1994).


\bibitem{arimondo76}
E. Arimondo and G. Orriols,
``Nonabsorbing atomic coherences by coherent two-photon transitions in a three-level optical pumping,''
Nuovo Cimento Lett. {\bf 17,} 333--338 (1976).


\bibitem{whitley76}
R. M. Whitley and C. R. Stroud, Jr.,
``Double optical resonance,''
Phys. Rev. A {\bf 14,} 1498--1513 (1976).


\bibitem{hunn95}
J. D. Hunn, S. P. Withrow, C. W. White and D. M. Hembree,
``Raman scattering from MeV-ion implanted diamond,''
Phys. Rev. B {\bf 52}, 8106--8111 (1995).


\bibitem{uzan_saguy95}
C. Uzan-Saguy, C. Cytermann, R. Brener, V. Richter,
M. Shaanan and R. Kalish,
``Damage threshold for ion-beam induced graphitization of
diamond,''
Appl. Phys. Lett. {\bf 67}, 1194--1196 (1995).


\bibitem{orwa99}
J. O. Orwa, K. W. Nugent, D. N. Jamieson, and S. Prawer,
``Raman investigation of damage caused by deep ion implantation
in diamond,''
Phys. Rev. B {\bf 62}, 5461--5472 (2000).


\bibitem{Zaitsev}
A. M. Zaitsev,
{\it Optical Properties of Diamond: A Data Handbook}
(Berlin: Springer, 2001).


\bibitem{davies76}
G. Davies and M.~F. Hamer,
``Optical studies of the 1.945 eV vibronic band in diamond,''
Proc. R. Soc. London Ser. A {\bf 348,} 285--298 (1976).


\bibitem{martin99}
J. P. D. Martin,
``Fine structure of excited 3E state in nitrogen-vacancy centre in diamond,''
J. Lumin. {\bf 81,} 237--247 (1999).


\bibitem{manson06}
N. B. Manson, J. P. Harrison, and M.J. Sellars,
``The nitrogen-vacancy center in diamond re-visited,''
preprint: \url{http://arxiv.org/abs/cond-mat/0601360}.


\bibitem{manson05}
N. B. Manson and J. P. Harrison,
``Photo-ionization of the nitrogen-vacancy center in diamond,''
Diamond \& Related Materials {\bf 14,} 1705--1710 (2005).


\bibitem{vanoort91}
E. van Oort, B. van der Kamp, R. Sitters, and M. Glasbeek,
``Microwave-induced line-narrowing of the N-V defect absorption in diamond,''
J. Lumin. {\bf 48 \& 49,} 803--806 (1991).


\bibitem{olivero2005}
P. Olivero, S. Rubanov, P. Reichart, B. Gibson, S. Huntington, J. Rabeau, A.~D. Greentree, J. Salzman, D. Moore, D.~N. Jamieson, and S. Prawer,
``Ion beam assisted lift-off technique for three-dimensional micromachining of free standing single-crystal diamond,''
Advanced Materials {\bf 17,} 2427--2430 (2005).

\end{thebibliography}
\end{document}